\documentclass[prd, twocolumn, nofootinbib]{revtex4}

\usepackage{epsfig} 

\newcommand{\beq}{\begin{equation}}
\newcommand{\eeq}{\end{equation}}
\newcommand{\beqa}{\begin{eqnarray}}
\newcommand{\eeqa}{\end{eqnarray}}
\newcommand{\om}{\Omega_m}

\def\la{\mathrel{\mathpalette\fun <}}
\def\ga{\mathrel{\mathpalette\fun >}}
\def\fun#1#2{\lower3.6pt\vbox{\baselineskip0pt\lineskip.9pt
  \ialign{$\mathsurround=0pt#1\hfil##\hfil$\crcr#2\crcr\sim\crcr}}}

\begin{document} 

\title{Dark Energy with Fine Redshift Sampling} 
\author{Eric V.\ Linder} 
\affiliation{Berkeley Lab, University of California, Berkeley, CA 94720} 

\date{\today} 

\begin{abstract} 
The cosmological constant and many other possible origins for acceleration 
of the cosmic expansion possess variations in the dark energy properties 
slow on the Hubble time scale.  Given that models with more rapid 
variation, or even phase transitions, are possible though, we examine 
the fineness in redshift with which cosmological probes can realistically 
be employed, and what constraints this could impose on dark energy 
behavior.  In particular, we discuss various aspects of baryon acoustic 
oscillations, and their use to measure the Hubble parameter $H(z)$.  We 
find that currently considered cosmological probes have an innate 
resolution no finer than $\Delta z\approx0.2-0.3$. 

\end{abstract} 

\maketitle 

\section{Introduction} \label{sec:intro} 

The origin of the acceleration of the cosmic expansion is unknown 
and currently observations give significant constraints on only 
broad properties 
of the physics, such as a time-averaged or constant effective equation 
of state ratio $w$.  To look for clues to the physics we seek to measure 
the dynamics with redshift, $w(z)$, but this requires next generation 
experiments with improved precision and systematics control, and a 
longer redshift range for leverage on cosmological parameter degeneracies. 

Characterizing the dark energy by a tilt $1+w$ and a running $w'=\dot w/H$ 
defines deviations from the cosmological constant value of $w=-1$ with no 
time variation, $\dot w=0$, and gives important distinctions between 
models or classes of physics (see, e.g., \cite{caldlin}).  Since the 
dark energy, whatever it is, lives in an expanding spacetime with 
characteristic expansion timescale, $H^{-1}=a/\dot a$, 
entering as a Hubble friction in the dark energy equations of motion, then 
dark energy variation may be slow, $w'\ll1$.  Models that exhibit rapid 
variation recently, after Hubble friction should have had a long time 
to influence them, tend to be highly fine tuned (see, e.g., 
\cite{pngb,paths}). 

Nevertheless, it is an interesting question to ask whether a rapid 
variation or phase transition in dark energy properties could be 
detected by cosmological probes by turning up the vernier, or fineness 
with redshift, of the measurements.  As we try to increase the redshift 
resolution, various effects thwart us: dependence of observables on 
single or double integrals of $w(z)$, broad redshift kernels of 
cosmological sensitivity, Nyquist and statistical limits on independent 
modes, and measurement systematics coherent over redshift. 

In \S\ref{sec:baohc} we explore the ``single integral'' probe of $H(z)$ 
from baryon acoustic oscillations and corrections to that picture. 
We discuss the cosmological sensitivity kernel of probes 
in \S\ref{sec:fine}, addressing the trade off between resolution and 
precision.  Important physical consequences of rapid time variation 
are mentioned 
in \S\ref{sec:rapid}.  We compare the ability of different baryon acoustic 
oscillation surveys to determine dark energy dynamics in \S\ref{sec:efold}, 
and sum up the results in \S\ref{sec:concl}.

\section{Measuring $H(z)$} \label{sec:baohc} 

Observables that depend on distances involve a double integral of 
the dynamics $w(z)$, increasing the difficulty of seeing a rapid 
variation.  Growth factors of mass density perturbations face much 
the same situation.  Measurements directly of the Hubble parameter 
$H(z)$ involve a single integral and so might offer more sensitivity 
to rapid variation.  The characteristic length scale of baryon 
acoustic oscillations (BAO), seen 
in the pattern of large scale structure clustering, in the radial, line 
of sight, direction is frequently phrased in terms of measuring $H(z)$ 
\cite{bg03,linbo,seoeis03}. 

The radial modes involve the ratio of the proper distance $r_p$ across 
a redshift interval $dz$ to the sound horizon scale $s$ at CMB last 
scattering, where 
\beq 
\int dr_p=\int_{z_c-\Delta}^{z_c+\Delta} \frac{dz}{(1+z)\,H(z)}, 
\eeq 
with $z_c$ the central redshift and bin width $2\Delta$. 
For thin redshift slices $H(z)$ generally varies little over the interval 
and the measurement of $I\equiv\int dr_p/s$ provides a good estimation of 
$H(z_c)$.  However we are specifically interested in the case when 
there {\it is\/} rapid variation, so it is worthwhile investigating 
the accuracy of this interpretation. 

The deviation in the true quantity $I$ from the approximation 
$\tilde I\equiv 2\Delta/[(1+z_c)\,H(z_c)\,s]$ often used to interpret 
the radial BAO as a $H(z)$, or single integral of $w(z)$, measurement, 
is quadratic in the bin width. 
The first order term vanishes from a Taylor expansion of 
the behavior of $H(z)$ about the central redshift, leaving 
\beq 
\frac{\delta I}{I}\approx 
\frac{(n+1)(n+2)}{6}\left(\frac{\Delta}{1+z_c}\right)^2, 
\eeq 
where $H(z)\sim (1+z)^n$ over the redshift interval.  
For the matter dominated case, $n=3/2$.  A typical value for $\Delta/(1+z_c)$ 
might be 1/16, e.g.\ for $\Delta=0.25$ at $z_c=3$, or $\Delta=0.1$ at 
$z_c=0.6$.  This would give a 0.6\% deviation $\delta I/I$.  For a LCDM 
cosmology, 
the deviation from a pure measurement of $H(z)$ varies from 0.2-0.8\%. 

A rapid transition, however, can give a correction first order in 
$\Delta/(1+z_c)$, because a Taylor expansion about the central 
redshift is invalid.  This holds as well for a phase transition that 
occurs within the redshift slice, above or below the central redshift. 
For example, a phase transition from $w=0$ behavior to $w=-1$ 
can create a deviation of 1.1\% of the 
BAO scale relative to the Hubble parameter $H(z_c)$ interpretation 
in a slice from $z=1.1-1.5$, 
and exceeds 6\% deviation for $z=0.4-0.6$.  Such a phase transition 
admittedly appears drastic, so we consider a more sedate model for 
describing a transition, the e-fold model \cite{lh05}. 

The e-fold model describes a transition from a high redshift value 
$w_p$ of the equation of state to an asymptotic future value 
$w_f=w_p-\Delta w$, occurring at scale factor $a_t=1/(1+z_t)$ with 
rapidity $\tau$: 
\beq 
w(a)=w_f+\frac{\Delta w}{1+(a/a_t)^{1/\tau}}. 
\eeq 
This has the virtue of a smooth, tunable transition and an analytic 
form for $H(z)$.  (One can also consider the kink model \cite{kink} 
though this lacks an analytic expression for $H(z)$.)  Taking a reasonably 
mild transition with $w_f=-1$, $w_p=-0.5$, $a_t=0.5$, and $\tau=0.2$ 
(i.e.\ $dw/d\ln a_t=-0.625$, a bit slower than the ``natural'' value 
$-1$) we find deviations of 0.7-0.9\% for slices of thickness 
$\Delta z=2\Delta=0.4$ at various redshifts in $z=1-2$. 

So in reality, the radial BAO scale measures an effective Hubble parameter 
\beq 
\bar H(z_c) \equiv \left[\int_{z_c-\Delta}^{z_c+\Delta} \frac{dz}{2\Delta} \frac{1+z_c}{(1+z)\,H(z)}\right]^{-1}
\eeq 
and is not formally a single integral probe of the dark energy dynamics. 
Because it is not actually measuring $H(z)$, the deviation between 
$\bar H$ and $H$, i.e.\ $\delta H/H=\delta I/I$, will bias the cosmology 
if it is not properly taken into account.  This bias will be small for 
experiments with precision weaker than 1\%.  
But even the LCDM case indicates that 
calling a radial BAO scale a measurement of $H(z)$ is inaccurate if 
the experiment aims at subpercent level precision. 

\begin{figure}[!hbt]
\begin{center} 
\psfig{file=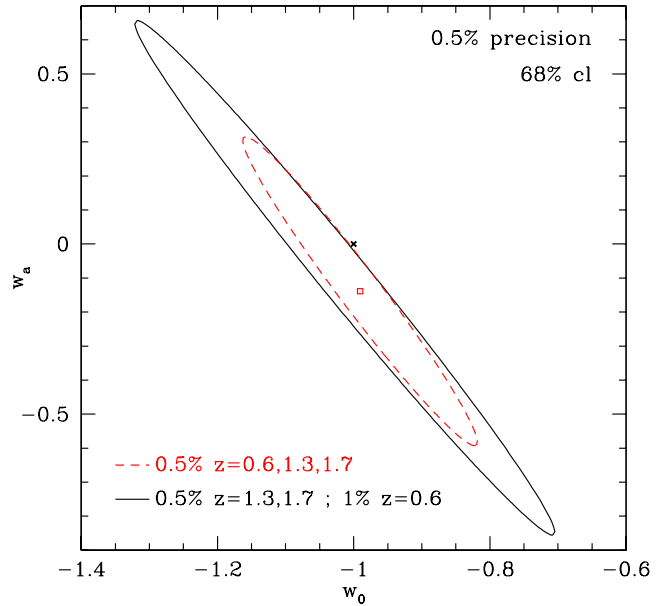,width=3.4in} 
\caption{Treating the baryon acoustic oscillation radial scale as 
a direct measure of the $H(z)$ biases the results from the true 
cosmology (here a cosmological constant, shown by the x).  Constraints 
here include a Planck CMB prior in addition to BAO radial and tangential 
scale measurements.  Much of the cosmology leverage comes from $z<1$, 
as shown by the weakening of the $1\sigma$ joint parameter uncertainty 
contour on relaxing the $z=0.6$ data precision from 0.5\% (dashed red, 
with best fit given by the square) to 1\% (solid black). 
}
\label{fig:baobias} 
\end{center} 
\end{figure}

Figure~\ref{fig:baobias} illustrates the effect of the misestimation 
of $H(z)$ on the cosmological model.  For a flat universe with 
dimensionless matter density $\om$ and dark energy equation of state 
$w(a)=w_0+w_a(1-a)$, the deviation from the true Hubble parameter 
can lead to an apparent disfavoring of the true, cosmological constant 
cosmology by more than $1\sigma$.  Here we consider a hypothetical 
experiment that 
obtains the BAO radial and tangential scales to 0.5\% precision (we 
take the tangential, or angular diameter distance, measurements to be 
unbiased) and combines them with 0.4\% precision on the distance to 
the CMB last scattering surface.  

Note that a BAO experiment covering 
only the ranges $z=1.1-1.5$ and $z=1.5-1.9$ (plus CMB) would not 
estimate the dark energy time variation to better than $\sigma(w_a)\approx1$, 
so we also add BAO radial and tangential information from an additional 
experiment in $z=0.5-0.7$.  This additional experiment actually carries 
much of the cosmological leverage, as opposed to the $z\approx1-2$ 
measurements, as seen by relaxing the lower redshift experiment's 
precision to 1\% (difference between solid and dashed curves).  In 
either case, the bias from treating the BAO radial scale as a measure 
of $H(z)$ must be addressed.

One can formally correct $\bar H$ to $H$, obtaining a true measure of $H$, 
only if one knows the cosmological model.  In this sense, interpreting 
radial BAO measurements in terms of the Hubble parameter runs into 
a systematic floor at the subpercent level.  

This is easily sidestepped, 
however, in that there is no need to claim measurement of $H$; one can 
simply carry out the analysis in terms of the radial distance $\int dr_p$, 
and propagate this through in the usual way to constraints on the 
cosmological parameters.  
This avoids bias, and only smears the parameter estimation by less than 
2\% relative to the case where $H$ had actually been obtained. 

One can also just assume a fiducial cosmological model to approximate 
the correction of $\bar H$ to $H$.  Specifically, we define 
\beq 
\bar H_{\rm cor}(z_c)=\bar H(z_c)+H_{\rm fid}(z_c)-\bar H_{\rm fid}(z_c), 
\eeq 
where fid denotes the fiducial model, e.g.\ LCDM with $\om=0.28$. 
Since the fiducial model is not necessarily the true cosmology, 
this still possesses bias but generally at a reduced level relative 
to no correction.  For example, in the 
e-fold case used above, the 0.7-0.9\% deviations become reduced to 
$\sim0.1$\%.  One could carry out an iterative correction process 
for further reduction. 

In practical terms, then, one does obtain values $H(z_i)$ for a set of 
redshift bins centered at $z_i$ (putting aside the exception of a model 
with a phase transition).  The limit on the fineness of 
the spacing of the $z_i$ is set by a variety of factors, including 
the ``Nyquist'' lower limit on the slice thickness by requiring a 
fundamental wavemode 
(determined by the comoving sound horizon scale of $\sim100 h^{-1}\,$Mpc) 
to fit across the redshift slice, 
\beq 
\Delta z>H\lambda_{\rm Nyq}=\frac{H}{H_0}\frac{2\times 100 h^{-1}\,{\rm Mpc}}
{3000 \,h^{-1}\,{\rm Mpc}}\approx 0.12-0.3 
\eeq 
where the last expression is for values of $H$ at $z=1$ and $z=3$ in 
a LCDM universe, 
and by the need for a large volume (hence substantial redshift thickness 
since the solid angle is limited by $4\pi$) to measure many modes to 
obtain strong statistical precision.  For percent level measurements, 
this last factor imposes \cite{seoeis,glazeblake} 
\beq 
\Delta z>0.2\,.
\eeq

\section{Fineness and Finesse} \label{sec:fine}

For the case of distance measures (and related probes), we can consider 
whether there is an innate limit to a fine-toothed cosmological comb for 
seeing rapid transition in dark energy dynamics, or whether increasingly 
finely sampled data can provide such a tool. 

The characteristic feature size of a variation is 
\beq 
\Delta\ln a\approx \Big|\frac{\Delta w}{dw/d\ln a_t}\Big|\,.
\label{eq:dlna} 
\eeq 
For appreciable transitions $|\Delta w|>0.2$ and non-extreme variations 
$w'\equiv dw/d\ln a<1$, we see that $\Delta\ln a>0.2$ or $\Delta z>0.2$. 
Thus very fine redshift sampling does not in itself enable further 
insights into the dynamics.  

\begin{figure}[!hbt]
\begin{center} 
\psfig{file=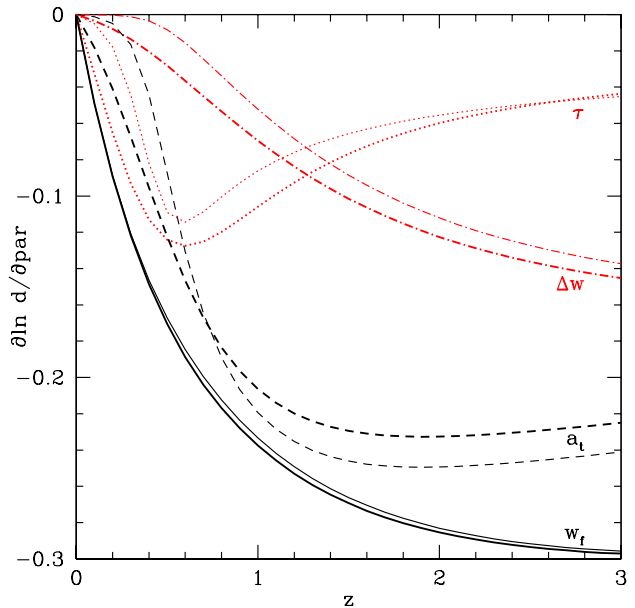,width=3.4in} 
\caption{The variation of distance observables with dark energy 
dynamics is smooth, with a broad kernel, even for rapid transitions. 
Here an e-fold transition from $w_p=-0.5$ to $w_f=-1$ takes place 
at $a_t=0.7$ ($z_t=0.43$), with $w'=-1$ (heavy curves) or $w'=-2.5$ 
(light curves). 
}
\label{fig:sens} 
\end{center} 
\end{figure}

Looking at the sensitivity functions $\partial\ln d/\partial p$ for the 
distance $d$ (whether this is luminosity distance or angular diameter 
distance, or the reduced distance $(\om h^2)^{1/2}d$ that BAO measure) 
and parameters $w_f$, $\Delta w$, $a_t$, 
$\tau$, we see from Fig.~\ref{fig:sens} that the dependences are 
fairly smooth in redshift.  Effectively, the kernel over which the 
observations feel the dark energy dynamics is broad, so fine mapping 
over redshift is not essential.  

For measures of 
$H(z)$, the characteristic redshift fineness for variation is somewhat 
enhanced, but still leads to $\Delta z>0.2$ for $w'<2$.  True phase 
transition behavior can avoid this bound, but as we saw 
in \S\ref{sec:baohc}, precision measures of $H$ through radial BAO 
require $\Delta z>0.2$ anyway.

For the e-fold model, $\Delta\ln a>4\tau$; see \S\ref{sec:rapid} for 
what happens if we try to make too rapid a transition ($\tau\ll1$). 
Note that the two parameters of most physical interest, the asymptotic 
future equation of state (hence determining the fate of the universe) 
$w_f$ and the equation of state change $\Delta w$ (or asymptotic past 
equation of state $w_p=w_f+\Delta w$) have very smooth variations, 
indicating broad sampling is sufficient.  Even the less interesting 
variables of the transition scale factor $a_t$ and breadth $\tau$ 
do not show fine structure with redshift (note Fig.~\ref{fig:sens} 
is for low $z_t$ and large $w'$, each acting to enhance the 
sharpness of features).  We can understand the 
sensitivity curves as follows: $\Delta w$, $a_t$, and $\tau$ have 
little sensitivity at redshifts between the present and the transition, 
while well after the transition the sensitivity levels off (for $\Delta w$) 
or slowly declines ($a_t$, $\tau$) as the equation of state becomes 
effectively averaged over.  For $w_f$ the sensitivity increases at low 
redshift (from zero due to the initial cosmology independent $d\sim z$ 
behavior), then gradually levels off, again due to effective averaging. 

The broad kernels seen here, in a model with an explicit transition, 
also exist in the principal component analysis of non-parametric 
$w(z)$.  The major principal components for supernovae, BAO, or 
weak lensing shear power spectrum all have widths $\Delta z\ga 0.3$, 
with weak lensing tending to have the broadest 
\cite{hutstar,lh05,pogos,hutpeir}.  
None of the major principal components act as a fine-toothed comb for 
dark energy.  While the weaker principal components do exhibit rapid 
variation in redshift, these components are poorly determined and 
the variation is oscillatory, so they do not really focus on specific 
dark energy behavior.  Localized, and less oscillatory, uncorrelated 
``square root'' principal components \cite{coohutpc} 
again show characteristic widths $\Delta z>0.2$. 

We now verify explicitly that fineness of redshift mapping, beyond 
$\Delta z\approx0.1$, does not improve determination of dark energy 
dynamics.  Consider distance measurements of some precision $P$ at $N$ 
redshifts evenly spaced in $z=0.1-1.7$ (we add a local anchor 
at $z=0.05$ as well) so the redshift fineness is $\delta z=1.6/N$ 
($\delta z$ is the sampling scale of the distance-redshift relation, 
not an error on redshift measurement).  As we increase $N$, giving 
finer mapping 
of the distance-redshift relation, or expansion history, the 
constraints on dark energy parameters improve.  However, this is not 
due to the fineness of the mapping, but to the additional 
statistics, i.e.\ the larger data set from increased sampling.  
Compensating for the statistics 
by reducing the ``finesse'', or precision on a measurement, i.e.\ 
keeping $PN^{-1/2}$ fixed, we find that the improvement arises 
essentially wholly from statistics.  That is, the dark energy 
constraints for the case of mapping with 
fineness $\delta z=0.1$ are within 2\% of those for mapping with 
$\delta z=0.05$ but $\sqrt{2}$ worse precision, and within 4\% for 
mapping with $\delta z=0.2$ but $\sqrt{2}$ better precision.  
Within these limits, the fineness of the experiment is matched by 
the finesse of the experiment. 

When the sampling becomes too broad, information {\it is\/} lost 
if the mapping misses the feature, and cannot be made up by $\sqrt{N}$ 
statistics; the cosmological leverage decreases.  As well, 
making the sampling sparser without continually scaling the precision 
worsens the cosmological constraints, while 
the statistical precision cannot be improved without limit due to 
systematic uncertainties imposing a floor. 

Conversely, finer sampling, which allows looser precision, leads to 
no gain because of the width of the cosmological kernel: reducing 
$\delta z$ to 0.01, while relaxing the precision by $\sqrt{10}$, 
gives constraints only 2-4\% weaker than the $\delta z=0.1$ case.  This 
seems to suggest an experiment capable of fine mapping, 
such as supernova distance measurements (but not radial BAO due to 
its redshift thickness requirements) might be employed in this 
``fine scan'' manner with a relaxed precision requirement.  An 
obstacle that arises to this approach is that systematics have 
not only an amplitude imposing a floor to precision but a redshift 
coherence length.  Sampling more finely than the coherence scale does 
not give independent measurements of the cosmology and so the fine 
sampling does not reduce the measurement uncertainty as fast as 
$N^{-1/2}$ and thus the constraints weaken.

\section{Rapid Transitions} \label{sec:rapid}

For fine features to survive in the cosmological expansion history, 
extreme dynamics of dark energy is required. 
We saw that the characteristic redshift width of the variations in 
observables is $\Delta z\gtrsim0.3$, only approaching somewhat smaller 
values when dealing with extraordinarily rapid transitions in dark energy 
and a probe capable of measuring $H(z)$, such as radial BAO.  As 
below Eq.~(\ref{eq:dlna}), the most optimistic scenario requires 
$|w'|>2$ or $|\dot w|>2H$. 

Apart from the practical difficulties discussed of making precision 
radial BAO measurements in thin redshift slices, and theoretical 
naturalness considerations of large $w'$ (and some observational 
limits on phase transition-like behavior, e.g.\ \cite{caldpark,early}), 
such rapid variation shakes 
the entire 
cosmological framework.  Most simply put, if the equation of state 
is to stay finite, say between 0 and -1, then a 
large $\dot w$ also requires a large $\ddot w$, i.e.\ a large velocity 
within a finite box requires a large acceleration too.  The quantity 
$\ddot w$ can also be phrased in terms of $w''$ by 
\beq 
\frac{\ddot w}{H^2}=w''+\frac{\dot H}{H^2}w'=w''-w'(1+q), 
\eeq 
where $q$ is the deceleration parameter.  (Note that for both the 
e-fold model and $w_a$ model, $w''\sim w'$.)  

However, rapid 
time variation in the dark energy equation of state is directly 
related to spatial variations in the dark energy, giving the field 
$\phi$ an effective mass $m=\sqrt{V_{,\phi\phi}}$, where the 
curvature of the potential 
\beq 
V_{,\phi\phi}/H^2\sim {\cal O}(w'^2,w''). 
\eeq 
(see \cite{paths,cald00} for the full expression).  If $m\gtrsim H$ 
then spatial inhomogeneities in the dark energy can form on subHubble 
scales -- i.e.\ it becomes a clustering component and affects large 
scale structure (see, e.g., \cite{beandore}).  Thus for rapid variations 
we have no guarantee that the primordial acoustic scale imprinted in 
large scale structure remains pristine (cf.\ \cite{dedeo}); the case 
when there might exist fine features in 
$H(z)$ is also the case when BAO may no longer function cleanly as 
a cosmological probe.

\section{Detecting a Transition} \label{sec:efold}

Given that an arbitrarily fine map of the expansion history is 
impractical, we consider how well we could detect rapid 
variation of the dark energy dynamics.  One problem with such 
behavior is the necessity for many parameters to describe the 
transition: the minimum as discussed in \cite{lh05,kink} is four 
equation of state variables, such as the asymptotic past and 
future values, the time of the transition, and its rapidity. 
However, even a combination of next generation data sets will be 
unable to accurately constrain all four parameters (plus the matter 
density and any other variables; we take a flat universe) -- 
the maximum number of tightly 
fit dark energy parameters is two \cite{lh05}. 

Therefore we must concentrate on the most interesting physics, 
holding the other parameters fixed, in order to obtain significant 
constraints.  Perhaps most interesting is $w_f$, the asymptotic 
future value of $w(z)$.  This plays a central role in determining 
the fate of the universe, e.g.\ if $w_f<-1$ one has classically a 
``Big Rip'' scenario \cite{bigrip}.  Moreover, the value $w_f=-1$ 
leads to an asymptotic deSitter spacetime, qualitatively different in 
its horizon properties from other cases \cite{weinberg}.  We might 
also like to know the amount by which the dark energy evolves, 
$\Delta w$, to check consistency with a cosmological constant, 
or equivalently know 
what its high redshift properties are -- e.g.\ did it track during the 
matter dominated era (e.g.\ \cite{wett,ratrap,tracker}), or was the 
early equation of state consistent with a geometrically induced value, 
such as $w(z\gg1)=-0.5$ for the DGP braneworld case (\cite{dgp,luess}; 
also see \cite{dvalitur}). 
The transition time and rapidity variables, $a_t$ and $\tau$ within 
the e-fold model, give a less direct view of the physics, and so, 
since we are forced to fix two variables, we hold them constant. 
This is not wholly satisfactory, but otherwise we cannot explore the 
impact of cosmological probe design on discriminating rapid transitions 
in dark energy.  (Note \cite{blanchard} addresses the complementary 
question of detectability of $a_t$ and $\tau$, fixing $w_f$ and 
$\Delta w$ -- their Eq.~2 is the e-fold model, with $\Gamma=1/(2\tau)$.) 

Generally, probes tied to the recent universe, such as supernova 
distances, do better at constraining $w_f$, and probes tied to the 
early universe, such as BAO scales, do better at constraining $\Delta w$. 
The crossover in advantage of one vs.\ the other is sensitive to the 
fiducial values assumed for the transition time and rapidity, so we 
cannot draw model independent conclusions about these two different 
techniques.  Instead, for a somewhat more controlled analysis we 
compare the BAO probe to itself by considering the effect of varying 
the redshift range of the data. 

We consider 1\% measurements of the radial and tangential acoustic 
scales (essentially $H(z)/(\om h^2)^{1/2}$ and $(\om h^2)^{1/2}d(z)$) 
for the cases $z\le1.0$ ($z=0.4-0.6$, $0.6-0.8$, $0.8-1.0$), 
$z\le1.2$ ($z\le1.0$ plus $1.0-1.2$), and $z<2$ ($z\le1.2$ plus 
$1.25-1.55$, $1.55-1.9$).  Note that measurements in the $z=1.2-1.9$ 
range are difficult from the ground, so we can examine whether the BAO 
technique presents a pressing need for space based observations. 

\begin{figure}[!hbt]
\begin{center} 
\psfig{file=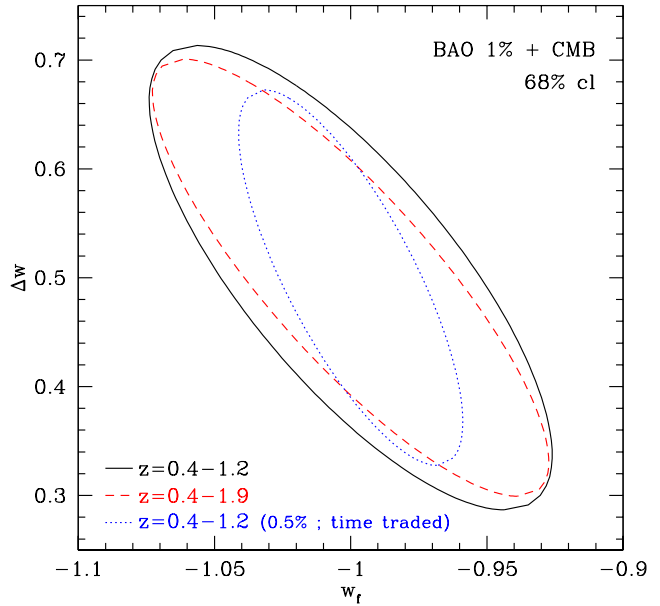,width=3.4in} 
\caption{BAO constraints on dark energy undergoing a rapid transition 
gain most of their leverage from data at $z\lesssim1$, due to the BAO 
scale being tied to the CMB at high redshift.  Trading the time/expense 
of $z>1.2$ BAO observations to improve the precision of $z<1.2$ 
information is much more efficacious. 
}
\label{fig:bzrg} 
\end{center} 
\end{figure}

Figure~\ref{fig:bzrg} shows the constraints on the dark energy 
variables $w_f$ and $\Delta w$ (marginalizing over $\om$ but fixing 
$a_t=0.5$ and $\tau=0.125$, i.e.\ $w'=-1$) for a fairly rapid transition 
in dark energy.  The asymptotic future value is determined to about 
5\% and the degree of evolution to $\sim30\%$.  It also illustrates 
that most cosmological leverage comes from the data at $z\la1.2$, 
accessible from the ground.  Adding the space observations in 
$z=1.2-1.9$ tightens the parameters constraints by less than 6\%. 
Reducing the range to $z\le1.0$ in turn weakens the constraints by 
9\% (but further narrowing of the redshift range rapidly diminishes 
leverage).

Despite half of the dark energy transition lying above $z=1$, the 
BAO technique 
is sufficiently powerful that observations at $z\la1$ provide most 
of the cosmological leverage.  This is due in large part to the BAO 
scale being tied to the CMB at high redshift, so in this sense 
$z<1$ has a greater lever arm.  The conclusion that the $z\le1.2$ 
data set suffices holds even for higher redshift transitions.  If 
the transition takes place at $z_t=2$, well above $z=1.2$, 
addition of $z=1.2-1.9$ data improves constraints by less than 
13\%.  Furthermore, recall this was for a very sharp transition, 
$w'=-1$; if we consider a transition at $z_t=2$ with $w'=-0.5$ 
(the equivalent of $w_a=1.5$ -- still rather rapid!), then the 
$z=1.2-1.9$ data deliver only a 3\% improvement.  This is good news, 
in that even cosmologies with rapid evolution in dark energy at 
$z>1$ do not require space based BAO observations; ground based, precision 
measurements at $z\le1.2$ suffice.

\section{Conclusions} \label{sec:concl}

If there is fine structure in the dark energy dynamics, i.e.\ rapid 
variation in $w(z)$, this will be difficult to detect by any 
currently considered cosmological probe.  
Cosmological sensitivities to variation have a finite kernel 
in redshift due to innate dependences of the observables, plus additional 
effects such as Nyquist limits of wavemodes in a redshift slice and 
coherence of systematics over redshift. 
(A hypothetical, idealized probe of arbitrary precision and no systematics 
can evade these conclusions.) 

The radial baryon acoustic oscillation scale effectively provides a 
measure of the Hubble parameter $H(z)$, though measurements aspiring 
to subpercent level accuracy must either carry out the analysis in 
terms of the finite proper distance interval or correct for bias. 
The corrected Hubble parameter is close to a single integral over the 
dark energy dynamics, but the fineness of resolution is restricted to 
$\Delta z\gtrsim 0.2$ by the Nyquist limit, the need for sufficient 
volume, and the innate cosmological kernel. 
Double integral measures such as tangential BAO and supernova distances 
have a similar resolution.  

For BAO, accurate experiments in the 
redshift range $z=0.4-1.2$, readily accessible from ground based 
observations, would provide the great majority of this probe's leverage,  
even for a transition occurring at $z\approx2$.  Supernova distances 
mapping $z=0-1.7$ have most discriminating power for transitions 
that are not extremely rapid or that occur at lower redshifts.  Rapid 
transitions raise theoretical complications in any probe 
utilizing large scale structure as one expects clustering of dark 
energy on subHubble scales, altering the matter power spectrum. 

An important consequence of the innate smoothing from the cosmology, as 
well as the other contributors such as systematics coherence, is that 
attempts to reconstruct the dark energy behavior utilizing derivatives 
of the data will say more about measurement noise than the 
intrinsically smoothed physics. 

The lack of a fine resolution probe in redshift, plus the increased 
number of parameters needed to describe a rapid transition in dark 
energy properties, mean that even next generation experiments will 
not be able to map dark energy behavior in great detail.  To draw 
the most information from the observations, survey design must 
concentrate on systematics control, including reducing coherence 
effects over redshift, and maximize the redshift baseline over the 
era when dark energy has significant influence.  In addition, we will 
be challenged to be clever enough to deduce the nature of dark energy 
from just a few, broad properties.

\section*{Acknowledgments} 

This work has been supported in part by the Director, Office of Science,
Department of Energy under grant DE-AC02-05CH11231.

\end{document}